\documentstyle[PASJadd,epsf]{PASJ95}

\newcommand{\vol}{51}
\newcommand{\no}{3}
\newcommand{\lett}{}
\newcommand{\spage}{1}
\newcommand{\rdate}{1998 June 24}
\newcommand{\adate}{1999 April 25}
\def \lta {\mathrel{\vcenter 
     {\hbox{$<$}\nointerlineskip\hbox{$\sim$}}}} 
\def \gta {\mathrel{\vcenter 
     {\hbox{$>$}\nointerlineskip\hbox{$\sim$}}}} 

\begin{document}
\setcounter{page}{1}

\title{Black-Hole X-Ray Transients:
The Effect of Irradiation on Time-Dependent Accretion 
Disk Structure}

\author{Soon-Wook {\sc Kim},\thanks{%
Korea Research Foundation Postdoctoral Fellow}
\\
{\it Department of Astronomy, Seoul National University,
                151-742, Seoul, Korea}
\\
{\it  E-mail: skim@astro.snu.ac.kr}
\\
J. Craig {\sc Wheeler},
\\
{\it Department of Astronomy, University of Texas at Austin,
         RLM 15.308, Austin, TX 78712-1083, U.S.A.}
\\
and 
\\
Shin {\sc Mineshige}
\\
{\it Department of Astronomy, Graduate School of Science, 
Kyoto University, Sakyo-ku, Kyoto 606-8502}
}


\abst{Some effects of irradiation on time-dependent accretion-disk models 
for black hole X-ray novae are presented.
     Two types of irradiation are considered: direct irradiation from 
the inner
hot disk and indirect irradiation as might be reflected by a corona or 
chromosphere above the disk.
     The shadowing effect of the time-dependent evolution of the
disk height and consequent blocking of the outer disk by the inner and middle
portions of the disk from the direct irradiation is included.
     The direct irradiation of the disk by inner layers where the 
soft X-ray flux is generated is found to have only a small effect
on the outer disk because of shadowing.  Mild indirect irradiation
that flattens, but otherwise does not affect the light curve substantially,
still has interesting non-linear effects on the structure of the disk 
as heating and cooling waves propagate.
   The irradiated disks do not always make simple transitions between the
hot and cold states, but can linger at intermediate temperatures or even 
return temporarily to the hot state, depending on the irradiation and the
activity in adjacent portions of the disk.}

\kword{Accretion disks --- Black holes --- Instabilities 
--- Stars: individual (A0620--00)--- Stars: X-ray }

\maketitle
\section{Introduction}

There is now broad recognition that the most common form of black-hole 
binary candidates in the Galaxy are transients (Chen et al. 1997).
The cause of the primary rise in outburst is most likely to be due to the disk 
thermal instability associated with the ionization of hydrogen and helium
(Cannizzo et al. 1982; Mineshige, Wheeler 1989).
The disk-instability model can also produce ultra-soft X-rays
($\lta$ keV) and a strong correlation between the optical and X-rays in the 
outburst evolution (Mineshige et al. 1990a).
There remain many interesting issues, however.
The original disk-instability model (without irradiation)
failed to produce a plateau in the $V$-magnitude like that
100--200 days after the peak in some systems.
Although the disk-instability model can to account for the 
beautiful exponential decays (Cannizzo et al. 1995), 
it has difficulty in producing the reflares (or secondary maxima) 
commonly observed in X-ray novae at 50--70 days after the main peaks.
These were suggested to be due to irradiation of the outer
portions of the disk (Mineshige, Wheeler 1989; Mineshige 1994).

We examine here some of the aspects of the effect
of irradiation on the propagation of heating and cooling waves
in time-dependent disks that are unstable to the ionization
thermal instability.  Preliminary versions of this work were
presented by Kim et al. (1994), Wheeler et al. (1996), Kim et al. (1996),
and Wheeler (1997, 1999).
A precursor to this work was that of Saito (1989), who was
the first to consider time-dependent, irradiated disks
with some simplifications (discussed later).
Because this is the first study of time-dependent irradiation
in black-hole disks, 
we concentrate on qualitative effects in systems like A0620--00.  
The model assumptions, including a prescription for
treating irradiation, are given in section 2.  The global effects of 
irradiation are presented in section 3 and the effects of irradiation
on the thermodynamics of the heating and cooling instability are 
given in section 4.  A discussion and conclusions are given in section 5.

\section{Disk-instability Model for Irradiated Accretion Disks}

\subsection{Irradiation}

Two types of irradiation are considered: direct irradiation from the  
innermost hot disk and irradiation that might be reflected by a corona or disk 
atmosphere or chromosphere above the disk
(Meyer, Meyer-Hofmeister 1994).
     The X-ray luminosity of the irradiation is given by 
\begin{equation}
\label{lx}
   L_{\rm X}(t)={\epsilon}{\dot M}_{\rm in}c^2,
\end{equation}
where the efficiency $\epsilon$ is 0.057 and 
${\dot M}_{\rm in}$ is the mass-accretion rate at the inner
edge of the disk $R_{\rm in}$, taken to be 3$r_g$ as 
for a Schwarzschild black hole. 
We adopt a simple model for indirect 
irradiation by assuming that the luminosity of equation (\ref{lx}) 
effectively arises from a point at the center of the disk ($R=0$).
The indirect irradiation flux is then given by
\begin{equation}
\label{fi}
    F_{\rm i}(R,t) = C_{\rm X} {{L_{\rm X}(t)}\over{4\pi R^2}},
\end{equation}
where $C_{\rm X}$ is a constant.
     More sophisticated models incorporating radiative transfer would give 
$C_{\rm X}$ = $C_{\rm X}(R, t)$ 
(see discussion in Tuchman et al. 1990), but such
time-dependent radiative transfer is too difficult at this time.

For direct irradiation, we follow the prescription of 
Fukue (1992; see also Kusaka et al. 1970).
This model assumes that the irradiation from a hot, geometrically
thin annulus near the inside of the disk
can be approximated by that from an infinitesimally
thin, filled, uniformly radiating surface centered on the black hole.  
For regions in the disk at distances much larger than
the radius of the annulus, this leads to the following expression
for the total flux incident on one surface of the disk:   
\begin{equation}
\label{fd}
     F_{\rm d}(R,t)=(1-A) {{L_X(t)}\over {4{\pi}R}}
             \cdot{d\over dR}\left[{{H(t)}\over R}\right]^2,
\end{equation}
where $A$ is the X-ray albedo.

In practice, this irradiation is computed to arise from a disk
with the radius of the inner zone and from a height above
the disk plane equal to the disk height, $H$, of the innermost
computed zone.  The latter is a function of time and is used
to compute the rays for direct irradiation of the outer disk.

Equation (\ref{fd}) is equivalent to that adopted by King et al. (1997,
see their equation 6).
It is smaller by a factor of $H/R$ than
the prescription used by van Paradijs (1996) for neutron-star 
systems.  For simplicity, we take the X-ray albedo to be $A$= 0.5.  
De Jong et al. (1996) have deduced an albedo 
of $\sim$ 90 \% for low-mass X-ray binaries, but
the specific choice of the albedo does not appreciably affect the disk
structure, since shadowing suppresses direct irradiation 
(see subsection 3.1). Equation (\ref{fd}) 
shows that direct irradiation is a function of the gradient 
of the disk height, and is hence very sensitive to small variations in the disk 
profile.  Analytic models assuming $d\ln H/d\ln R$ = constant may exaggerate
the irradiation by minimizing the shadowing that may occur even
in the steady state.  They are certainly not adequate for time-dependent disks.
The height of the region from which the irradiation arises varies
with time and the disk height profile varies with both the irradiation 
and time.  
This makes the shadowing of the outer disk a complex, time-dependent 
phenomenon. We assume that no direct irradiation flux is added to 
the shadowed portions of the disk, while the indirect irradiation heats 
the whole disk in accord with equation (\ref{fd}).

\subsection{Basic Equations and Boundary Conditions}

Once ${\dot M}_{\rm in}$, 
and hence $F_{\rm i}(R,t)$ and $F_{\rm d}(R,t)$, are determined, the
irradiated flux is added to the flux generated internally in the disk by
viscous heating through an implicit numerical method, and other 
physical variables are computed in each time step
(see Tuchman et al. 1990 for details).
A modification of the outer boundary condition by irradiation affects
the whole vertical structure of the disk, including the mid-plane
values (Tuchman et. al. 1990).  This is a reasonable first approximation,
but could be improved by including more sophisticated 
radiative transfer effects.
In this formulation, the irradiation fluxes are 
parameterized by two constants, $C_{\rm X}$ and $A$, 
which are, by assumption, independent of the radius and time.

The basic equations of the time-dependent thermally unstable 
accretion-disk model (for details, see Kim et al. 1992 and references therein) 
are the mass conservation,
\begin{equation}
\label{mass}
  2{\pi}R{{{\partial}{\Sigma}}\over{{\partial}t}}
       ={{{\partial}{\dot M}}\over{{\partial}R}} +2{\pi}R S_0(R,t),
\end{equation}
the angular momentum conservation;
\begin{equation}
\label{angmom}
  2\pi R{\partial(\Sigma\ell)\over\partial t}
       ={\partial(\dot M\ell)\over\partial R}
        +2{\pi}R{S_0(R,t){\ell_{\rm K}}}
        -{{\partial}\over{{\partial}R}}({2{\pi}R^2}W),
\end{equation}
and the vertically averaged energy equation;
\begin{equation}
\label{energy}
  {{{C_{\rm p}}{\Sigma}}\over 2}\left({{\partial}\over{\partial t}}
       +{v_R}{{{\partial}}\over{{\partial}R}}\right){T_{\rm c}}
       =Q^+_{\rm vis}+F_{\rm d}+F_{\rm i}
        - Q^-_{\rm rad}+Q_{\rm dif},
\end{equation}
where we use cylindrical coordinates ($R, \phi, z$),
with the origin at the black hole and the $z$-axis perpendicular
to the plane of the disk ($z=0$). Also,
$\Sigma$ is the surface density, $S_0(R,t)$ is the mass
source function defined by
\begin{equation}
  {\dot M}_{\rm T}(t)={\int}2{\pi}RS_0(R,t)dR,
\end{equation}
with ${\dot M}_{\rm T}$ being 
the mass transfer rate from the companion star, 
$\ell=(GMr)^{1/2}$ is the specific angular momentum of matter in the disk,
$\ell_{\rm K}$ is the specific angular momentum of the stream 
from a companion star, $M$ is the mass of black hole, 
$W$ $(=\int\alpha Pdz)$ is the integrated viscous stress (with 
$\alpha$ and $P$ being the viscosity parameter and pressure, respectively),
$T_{\rm c}$ is the central temperature of the disk,
$v_R={-{\dot M}}/{2{\pi}R{\Sigma}}$ is the radial velocity of
the matter,
$C_{\rm p}$ is the specific heat with constant pressure,
and ${\Omega}_{\rm K}$ is the Keplerian angular frequency.

There are several terms on the right-hand side of equation (\ref{energy}).
The disk heating in the non-irradiated disk 
is given by the viscous heating;
\begin{equation}
\label{qvis}
     Q^{+}_{\rm vis}~=~{3\over 4}W{\Omega}.
\end{equation}
     The external-heating terms due to indirect and direct irradiation
of the disk, $F_{\rm i}$ and $F_{\rm d}$, 
are given by equations (\ref{fi}) and (\ref{fd}).
In the present time-dependent study,
  we included irradiation as heating in the vertically averaged structure,
  and did not solve the vertical structure of the disk with
  irradiation input from above the surface (cf. Tuchman et al. 1990).
The radiative cooling is given by 
\begin{equation}
\label{qrad}
    Q^{-}_{\rm rad} = {\sigma}T_{\rm eff}^4,
\end{equation}
where $\sigma$ is the Stefan--Boltzmann constant.
Finally, $Q_{\rm dif}$ is the heat-diffusion term, defined by
\begin{equation}
Q_{\rm dif}
   ={{\nu}_{\rm th}}{{C_{\rm p}\Sigma}\over {2R}}{\partial\over{\partial R}}
     \left(R{{{\partial}{T_{\rm c}}}\over{{\partial}R}}\right),
\end{equation}
and ${\nu}_{\rm th}(=~2W/{\Omega}{\Sigma})$ is the thermal diffusivity.
Here, we use the $\alpha$ formula with the same $\alpha$ as the viscosity
(see below).

We employ the following viscosity parameter ($\alpha$) prescription in a
time-dependent manner (Mineshige, Wheeler 1989):
\begin{equation}
\label{alpha}
   {\alpha}(R,t)~=~
   {\alpha}_0~\left[{{H(R,t)}\over R}\right]^N,
\end{equation}
where ${\alpha}_0$ is a constant, and $R$ and $H$ are the disk radius 
and height.
     From the hydrostatic relation, $H$ = $C_{\rm s}/{\Omega}_{\rm K}$, 
where $C_{\rm s}$ is the sound speed,
we can rewrite equation (\ref{alpha}) as
\begin{equation}
   \alpha(R,t)~=~
   \alpha_0~{\left({{{\cal R}\over \mu}{T_{\rm c}}
                    ~v_\phi^2}\right)}^{N\over 2}, 
\end{equation}
where $\cal R$ is the gas constant, $\mu$ is the mean molecular weight, and
$v_{\phi}(R)$ = $R~{\Omega}_{\rm K}(R)$, the azimuthal velocity in the
Keplerian disk.
     This implies that $\alpha$ is a function of the mid-plane or central
temperature: $\alpha$ $\propto$ ${T_{\rm c}(t)}^{N/2}$.

The boundary conditions are
\begin{equation}
  {\partial T_{\rm c}\over\partial r} = 0   \quad {\rm and}\quad
  {\dot M} = {\dot M}_{\rm T}
\end{equation}
at the outer edge and
\begin{equation}
   T_{\rm c} = W = 0
\end{equation}
at the inner edge of the disk, where 
the mass-transfer rate from the companion star is
taken to be constant in any particular model.
Relativistic effects are not taken into account.

\subsection{Methods of Calculations}

The radially-dependent equations, (\ref{mass})--(\ref{energy}),
are solved with the equations obtained
by integration of the vertical structure of disks,
\begin{equation}
   f=f(T_{\rm c},\Sigma,R),
\end{equation}
where the function $f$ describes the disk parameters,
 such as the flux, $Q^-_{\rm rad}$, 
viscous heating, $Q^+_{\rm vis}$ (or $W$), 
disk height, $H$, central (mid-plane) density, $\rho_{\rm c}$, 
and optical depth, $\tau$.

\begin{fv}{1}  
{20pc}  
{Opacities from Alexander et al. (1993), Cox and Stewart (1970), and
Cox and Tabor (1976) for $\log\rho = -6.0$ to $-8.0$ at intervals of
$\Delta\log\rho$ = 0.5.}
\end{fv}
We have taken the opacity from Alexander et al. (1983) for low 
temperatures ($\log T_{\rm c}$ $\lta$ 4.0), and Cox and Stewart (1970) and 
Cox and Tabor (1976) for high temperatures 
($\log T_{\rm c}$ $\gta$ 4.0) with mass
fractions $X_{\rm H}$ = 0.71 and $X_{\rm He}$ = 0.27.
The opacity for the intermediate temperature region is interpolated.
The resultant opacity in the low-density region appropriate to our models
in the outer disk is presented in figure 1, where the dominant effects are
designated.  We plot the opacity for 
$\log\rho_{\rm c}$ = $-$6 to $-$8.  The disk regions in which
we are interested  have a density distribution
of $\log {\rho}_{\rm c}$ = $-$6.2 to $-$7.7.
The range, $\log T_{\rm c}$ $\sim$ 3.25--3.55, is the region dominated by 
molecular opacity.
As shown in figure 1, the effect of molecules diminishes at 
$\log T_{\rm c}$ $\gta$ 3.5, depending on the density.
At $\log T_{\rm c}$ $\gta$ 3.7, ${\rm H^{-}}$ first becomes important.
We define the domain between these two temperature
regions as an intermediate state.
The partially ionized region can reach well above $10^4$ K,
depending on the density.
In terms of the effective temperature, the hot state is roughly
$\log T_{\rm eff}$ $\gta$ $10^{3.85}$.

Two regimes where changes in the physical conditions can affect the 
stability of the disk have
been discussed in  previous studies.
One is the partially ionized, intermediate warm state centered at 
$\log T_{\rm eff}$ $\sim$ 3.8, or, equivalently, 
$\log T_{\rm c}$ $\sim$ 3.7--4.1
(Mineshige, Osaki 1985; Mineshige 1988; Kim et al. 1992).
The other regime occurs in the temperature region centered at
$\log T_{\rm eff}$ $\sim$ 3.4 
(or $\log T_{\rm c}$ $\sim$ 3.4$-$3.7) in a marginally optically
thin region (Cannizzo et al. 1982; Mineshige, Osaki 1983) dominated
by molecular opacity, such as ${\rm {H_2O}}$ (Cannizzo, Wheeler 1984).  
These regions of variable opacity can lead to ``metastable" or ``stagnation"
states in the evolution between the hot and cold stable states of the disk
(discussed in subsection 4.2).
Ionization stagnation
is explicitly displayed in models for the outburst rise by Mineshige 
(1988) and Kim et al. (1992).
The possible significance of the molecular opacity to the
disk-instability model was raised by
Cannizzo and Wheeler (1984), but no explicit presentation has been provided.

The basic implicit numerical code used to compute time-dependent 
disk-instability models is that given by Mineshige (1986; see also
Mineshige, Wheeler 1989; Kim et al. 1992).  
The present disk-instability models have been improved 
compared to those presented by Mineshige and Wheeler (1989), in which an 
unreasonably small disk size ($3.16{\times}10^{10}$ cm) 
was adopted so as to avoid 
numerical instabilities and in which irradiation is ignored.
These models represent an implicit computation of
21 radial zones separated by radial mesh points at fixed spacings of 
$\log {\delta}R \sim$ 0.22,
from $\log R_{\rm in}$ = 6.5 to $\log R_{\rm out}$ = 11.0.
This zoning is adequate to reproduce 
the qualitative features of the expected outburst light curves
for our adopted prescription for the viscosity parameter (see below).
 It is clearly not sufficient to provide a 
fully satisfactory model of a time-dependent irradiated disk.
These models represent the first study of both the radial and
time-dependent nature of irradiated disks around black holes
where the central star cannot provide a source of flux.  
They are presented as a step beyond the single
irradiated annuli studied by Tuchman et al. (1990)
and Mineshige et al. (1990b) and as a step toward
the ultimate solution of time-dependent irradiated disks.  
They should not be viewed as full disk models, but as collections 
of coupled annuli that, nevertheless, give some qualitative insight 
into the physics of irradiated, time-dependent accretion disks. 

\subsection{Model Parameters}

We choose a 4$M_{\odot}$ black hole with a companion of 0.27$M_{\odot}$ as
being representative of A0620--00 (Marsh et al. 1994 and references therein).
This choice, with the observed orbital period of 7.75 hr, 
gives the following
disk model parameters: $R_{\rm in}$ = $3.6{\times}10^6$ cm,
binary separation $2.24{\times}10^{11}$ cm,
Roche lobe radius of the primary system $1.83{\times}10^{11}$cm and
the disk radius $10^{11}$cm (Frank et al. 1992; Paczy{\'n}ski 1977).
Both the inner and outer boundaries of the disk are held fixed.
We tested a range of values of $R_{\rm out}$, 
and found that the qualitative behavior is similar for outer-disk
radii in the range
$5{\times}10^{10}$cm $<$ $R_{\rm out}$ $\lta$ $1.3{\times}10^{11}$cm,
the tidal radius of the disk.
We choose the mass-transfer rate from the companion star to be
${\dot M}_{\rm T}$ = $3.16{\times}10^{16}$g~s$^{-1}$. 
This is larger than that given by  McClintock and Remillard (1986), 
but within the observational uncertainty (see the discussion 
in McClintock et al. 1995).
The transferred mass from the companion is initially accumulated 
in the outermost disk ($6{\times}10^{10}-10^{11}$ cm)
and then diffuses and accretes inward.

The outburst recurrence time in these models is about 3--4 years.  
This is too short for most observed systems, but does not affect
the qualitative thermodynamics of the irradiation, which is
the current focus.  The recurrence time would be longer if the
transfer rate were smaller (see also King et al. 1997).
For this relatively short recurrence time, the inner disk does not
have time to return entirely to the cool state.  The mass-flow
rate in the inner disk, and hence the generated radiation, does
drop by many orders of magnitude (about 7) from the peak, which
is sufficient to capture the essence of the variable production
of irradiation and how it reacts with the outer disk during
the phases of heating and cooling.

Given the limitation of the small number of zones, 
we have checked the basic behavior of the code by computing the speed
of propagation of the cooling wave.  
Cannizzo et al. (1995)
express the speed of the cooling front as
\begin{equation}
     V_{\rm front}~=~{2\over 3}\alpha_0 \left( {H\over R} \right)^{N+1}
      {C_{\rm s}\over w}R,     
\end{equation}
where $H$~=~$C_{\rm s}{\Omega}$ is the vertical scale height at radius $R$,
C$_{\rm s}$ is the sound speed, $\Omega$ is the Keplerian angular velocity,
$w$ is the width of the cooling front and the viscosity 
parameter is given by $\alpha$ = $\alpha_0(H/R)^N$.
The results of our numerical models are in reasonable
agreement with this expression for our assumed 
$\alpha$ prescription and logarithmic zoning.
In the first 40 days after the maximum,
the velocity of the cooling wave, $V_{\rm front}$, is almost the same in
both the irradiated and the non-irradiated models, but slightly slower
in the irradiated case.  At later times, the cooling-wave
speed in the irradiated model is somewhat faster than that for the
non-irradiated case, although both become very slow, less than
$10^3$ cm s$^{-1}$ after 100 days.
The overall shapes of the velocity profiles are rather similar in
both models.

Although a certain range has been explored, 
we have not undertaken a systematic parameter study of these
computationally expensive irradiated models.
Varying $C_{\rm X}$, ${\alpha}_0$, and $N$ gives different slopes 
of the light curves in the decay and different intervals between 
the peaks of the optical activity and the
inner mass flow rate.
If the indirect irradiation efficiency parameter,
$C_{\rm X}$, is sufficiently large, $C_{\rm X}\sim10^{-2}$, 
the models will be heated to a permanently hot steady state
(see van Paradijs 1996).
To approximately reproduce the observed light curve of A0620--00, 
we adopt $N=2$, 
${\alpha}_0$ = $10^3$, and $C_{\rm X}$ = $1.85{\times}10^{-4}$. 
For the chosen parameters, the outbursts in these models always begin 
in the outer disk.  
This determines the systematics of the overall outburst.
In particular, for systems with a fast rise and exponential
decline, the optical outburst is predicted to arise well before the 
increase of the mass flow rate in the inner disk that might be associated
with harder flux. 

We have computed several repeated outbursts in each model to achieve
an approximate steady state.
Our current models with irradiation
show an outburst cycle of $\sim$ 4 yr.  
We did not attempt to reproduce longer recurrence times in this study,
since our main current consideration concerned the effects of irradiation on the
outburst, itself. 
In the disk-instability model, the recurrence time scale is related to
$\alpha$, ${\dot M}_{\rm T}$, $R_{\rm out}$ and the masses of 
the binary components, $M_1$ and $M_2$
(e.g., Cannizzo et al. 1988; Mineshige, Wood 1989).
In principle, the easiest way to achieve a longer
recurrence time would be to decrease ${\dot M}_{\rm T}$.  While
the value we have chosen here,
${\dot M}_{\rm T}$ = $3.16{\times}10^{16}$g~s$^{-1}$,
is plausible, it could be a factor of 10 smaller and still
agrees with the observational constraints on, e.g. A0620--00.

\section{Outburst Evolution}

\subsection{Light Curves}

\begin{fv}{2}  
{20pc} 
{Model light curves on the decline from the maximum
along with the optical observations of A0620--00
(Webbink 1978 and private communication).
The model optical maximum is defined as day zero.
The data for A0620--00 is plotted 1.5 mag brighter than observed
for clarity of presentation and their phase is set by
assuming coincidence of the observed soft X-ray peak and
the model optical maximum (see text).
The model with both direct and indirect irradiation fits
the observed optical light curve of A0620--00 with
an assumed inclination of 60$^{\circ}$ and a distance of 400 pc.}
\end{fv}

The optical light curves are computed by assuming that each
disk annulus radiates as a black body at temperature $T_{\rm eff}(R,t)$
and summing the emission in the $V$ band over the annuli.
The resultant model optical light curves are presented in figure 2
together with observations of A0620--00.
(The data for A0620--00 is plotted 1.5 mag brighter than the observed value
for clarity of presentation.)
The model with both direct and indirect irradiation 
gives an optical light curve with a fast rise, a maximum,
and a nearly exponential decline.
As can be seen from figure 2, the direct
irradiation has virtually no effect on the light curve.  The
modest indirect irradiation we have invoked gives a slight, but
noticeable flattening of the light curve.  More severe 
indirect irradiation would cause this flattening to be more
extreme.  For the irradiation parameter $C_{\rm X}\sim10^{-2}$, the disk
would be stabilized.  The nature of this irradiation stabilization
will be explored in future work.

Note that our calculations never run into a radiation pressure-dominated
regime nor an optically thin one,
probably because the peak luminosity remains a slightly below that at which
the radiation pressure dominates over the gas pressure. 

\subsection{ Mass Flow}

\begin{fv}{3}  
{20pc} 
{Time-dependent evolution of the innermost accretion rate
(${\dot M}_{\rm in}$) for cases
with no irradiation (short dash), direct irradiation only (long dash),
indirect irradiation only (dot-dash), and both direct and
indirect irradiation (solid line).}
\end{fv}

The model accretion rate through the inner boundary
($R_{\rm in}$) of the disk is presented in figure 3.
Note that since the mass flow rate depends sensitively on $R_1$ as 
$\dot M\propto R_{\rm in}^{2.6}$ and since
$R_{\rm in}$ is much larger than that of the marginally stable
 circular orbit ($3r_{\rm g}$) in our calculations, due to the crude zoning, 
our quiescent mass-flow rate is much larger than that given by
the fine-mesh calculations of Ludwig et al. (1994).  However,
the qualitative behavior should be adequately captured by these models.

The indirect irradiation leads to a lower ${\dot M}_{\rm in}$ in 
quiescence due to a higher depletion of the disk mass during outburst.
The mass-transfer rate through the inner edge of the disk
determines one component of the soft X-ray flux.
In the current models, ${\dot M}_{\rm in}$ does not begin to rise until
13.6 d  before the primary optical peak, over two weeks since the start 
of the optical display.  
The peak in the mass flow rate occurs about 20 d after the primary 
optical maximum and about 30 d after the first rise in the inner
mass-flow rate.
Qualitatively, these models predict that the optical flux from 
the outer disk should rise before any activity from the inner disk, 
soft X-rays from the inner optically thick disk or hard or 
soft X-rays from any coronal emission.
This is consistent with the results of multi-wavelength observations
of Nova Muscae (see Lund 1993).
A prediction of the absolute times of peak emission in any band is more
uncertain in the absence of reliable models for the time-dependent 
power-law emission.

\subsection{ The Shadow Effect}

The non-monotonic distribution of the disk height 
causes the middle portions of the disk to block the flux
of direct radiation emitted from near the inner edge of the disk, and
thus prevent the direct irradiation of outer parts of the disk.
The shadowing of the outer disk is determined by the radial distribution of
the opening angle, $\tan\theta=H/R$, assuming the radiation to arise
from an inner region of small $H$ and $R$.  A region with a given value
of $H/R$ will thus be shadowed by a region at smaller $R$ with larger $H/R$.
Outer regions with small $H/R$ in the wake of an inward-propagating
cooling wave will clearly be in the shadow of inner, hot regions.

\begin{fv}{4}  
{20pc}
{Shadow effect for the direct irradiation.
The hatched part in the figure signifies where a given zone
at a given time is in the shadow
of some interior zones that are exposed to direct irradiation.}
\end{fv}

Some systematics of the shadowing process are illustrated in figure 4.
The specific details reported here may be subject to the effects 
of finite zoning.  The qualitatively significant results arise 
from comparing models with and without irradiation.           
     
Figure 4 shows that the innermost disk ($\lta$ $2{\times}10^7$ cm) is always 
exposed to the direct irradiation.
Because the outburst is initiated about 25 d prior to the optical peak, more 
extensive portions of the disk 
are exposed to direct irradiation.
After maximum light, the outer disk ($>$ $2{\times}10^8$ cm) 
is essentially totally blocked from direct irradiation 
by the inner disk.
With the shadowing effect, therefore, the existence of a mechanism which
can reflect the radiation from the inner region is important in the
current models if there are to be any effects of irradiation in the 
outer disk.

\subsection{ Amplitude of Disk Irradiation}

\begin{fv}{5}  
{22pc} 
{Contribution of irradiation in the disk evolution.
For illustration, we choose four zones: 5 (short dash), 16 (dot),
18 (dash dot), and zone 20 (solid line).
The contribution of both direct and indirect irradiation is plotted
with respect to the total heating in the upper panel.
In the lower panel, we present the ratio of direct to indirect irradiation.
The direct irradiation shuts off its effect about 20 d
after outburst maximum (day zero) due to shadowing (see figure 4).}
\end{fv}

The disk heating in the non-irradiated disk-instability
model is given by the viscous heating, $Q^+_{\rm vis}$ 
[see equation (\ref{qvis})].
In the presence of irradiation, the total heating is described by 
\begin{equation}
     Q^{+}_{\rm tot}~=~Q^{+}_{\rm vis}~+~F_{\rm i}~+~F_{\rm d}.
\end{equation}
The presence of irradiation affects the disk structure, and hence
the contribution from $Q^{+}_{\rm vis}$, itself, is altered.  
The resultant contribution of the irradiation to
the total disk heating is presented in figure 5
for an inner, a middle, and an outer region 
with the model optical maximum again being defined as day zero.
In the inner region ($\lta10^8$ cm), the irradiation is 
negligible throughout the entire evolution.  The direct component is
about 10 \% of the indirect component in the inner region, 
but the total irradiative heating does
not contribute substantially compared to the viscous heating.
In the middle of the disk ($\sim10^{10}$ cm), 
the contribution of the irradiation
is small, but finite.   Direct irradiation is appreciable 
only during the early decay, and is truncated by 20 d due to shadowing.
In the outer disk ($\sim6\times10^{10}$ cm), 
the contribution of the irradiation becomes important.
Around maximum, the viscous heating dominates the irradiation. 
After the cooling wave propagates into the outer region, and this
region cools, the irradiation heating, due entirely to indirect
irradiation because of shadowing, dominates over the viscous heating. 
This is shown by the rapid increase in the 
ratio of irradiation heating to the total heating about 5 d after the 
maximum in figure 5. 
The indirect irradiation heating then decreases along with 
${\dot M}_{\rm in}$ as the cooling wave propagates inward.
The behavior of the outer regions under the effects of 
radiation is more complicated than reflected in figure 5.

\subsection{ Global Effects of Irradiation on Decay}

\begin{fv}{6}
{22pc}
{Time-dependent evolution of the disk mid-plane
temperature during the early decay phase (upper panel) and the
corresponding states of the gas (lower panel) for the model
with no irradiation.
``HOT" denotes the hot thermal equilibrium states
and ``MOLECULE" represents the molecular stagnation stage
(see text and figure 1).}
\end{fv}
\begin{fv}{7}
{22pc}
{Time-dependent evolution of the disk mid-plane
temperature during the early decay phase (upper panel) and the
corresponding states of the gas (lower panel) for the model
with both direct and indirect irradiation.
``HOT" denotes the hot thermal equilibrium states
``Partial Ionization" 
means that the disk is in the metastable ionization stagnation stage,
and ``MOLECULE" (or ``M") represents the molecular stagnation stage.
``Intermediate State" (or ``i") denotes the intermediate state between
the molecular and partially ionized region (see text and figure 1).}
\end{fv}

The influence of mild irradiation is not restricted to a small modification
of the light curves.  Rather, it causes crucial changes in
the local disk structure, as illustrated in figures 6 and 7.
The upper panel of figure 6
presents the time-dependent evolution of the disk central
temperature in the model {\it without~irradiation}.
This model reproduces the typical decay phase, as shown in numerous
other studies of the disk instability.
As the cooling wave propagates from the outer to inner radii, the disk makes
the transition from the hot to cool state from larger
to smaller radii.
Note the sharp edges of the rapid transition from the hot to
cool state, which represent the cooling wave propagation.
The lower panel of figure 6 shows the opacity domains through
which the disk evolves.  It is basically only in the hot, ionized
or cool, molecular states with very rapid transitions between them.
This is because
in the non-irradiated model, or in the model with only direct 
irradiation and shadowing, the stagnation phase is absent.

Figure 7 displays the time-dependent disk evolution 
in the decay phase for the irradiated disk-instability model 
{\it with~both~direct~and~indirect~irradiation}.
The upper panel of figure 7 shows the evolution of the disk central
temperature, $T_{\rm c}$.
The squares denote the time at which the direct irradiation is 
terminated at a given radius as it falls into the shadow of interior regions.
With irradiation, the evolution is
much more complex, as illustrated in figure 7.
At 14.2 d, the region at $\sim4\times10^{10}$ cm
reaches the critical surface density for a
thermal instability and begins the rapid cooling that characterizes
the propagating cooling front of figure 6.  
The same transition
is seen at $\sim2\times10^{10}$ cm after 41.6 d and 
at $\sim1\times10^{10}$ cm after 88 d.
Rather than dropping directly
to the cool state, however, these regions linger in an intermediate
``metastable" stagnation state.  This is obviously due to the
effects of irradiation, since this behavior is absent in
figure 6, the effects, however, are rather subtle and involve global
coupling in the outer regions.

Throughout the disk evolution from quiescence to outburst, the outermost
disk (around $10^{11}$ cm) stays in the cold, neutral state 
at $\log T_{\rm eff}$ $\lta$ 3.4, or $\log T_{\rm c}$ $\lta$ 3.55.
After maximum light, the cooling wave is initiated 
around $6\times 10^{10}$ cm.
The resulting outward diffusion of the surface density results in a small 
increase in the temperature at larger radii.
The molecular stagnation (discussed in subsection 4.2) and consequent 
higher temperature of the outer regions 
due to the indirect irradiation affects the behavior of the inner regions.  
The regions just interior to the molecular-stagnation
region can not simply drop
into the cool state, because there is now some resistance to
the outward diffusion of matter that is attendant to the 
non-irradiated cooling wave.  
Instead, the decline in temperature
after the onset of the cooling instability is halted
in the nearly constant temperature intermediate ionization stagnation
state, as shown in figure 7.

After the outburst maximum, the X-ray novae have
exhibited additional features
in the decay, such as  ``reflares" 50--80 d after the maxima, the
``second maxima" a few hundred days later,
and subsequent ``mini-outbursts" before returning to quiescence.
The reflare is common in many X-ray novae throughout a variety of
wavelengths: optical (A0620--00), UV (Nova Muscae 1991, GRO J0422+32) and
soft X-rays (A0620--00, Nova Muscae 1991, GRO J0422+32).
As shown in figure 7, one portion of the cooling disk undergoes
a heating instability that takes it back to the hot, ionized state,
thus producing a transient increase in the light curve (see figure 2).
This might be associated with the reflare, since the optical feature is
consistent with Kuulkers (1998), who summarized the data
for the reflare in A0620--00 in both optical and X-rays (with some delay). 
It is not clear whether the optical feature in our models has some 
physical basis or whether it is an artifact of the crude zoning.
Our models do not provide any modulation of the inner mass flow and
hence of the soft X-rays at the time of the reflare.  Further studies
of these issues are needed.

\section{Irradiation and Thermodynamics}

\subsection{The Limit Cycle}

\begin{Fv}{8}  
{33pc}
{Thermal limit cycles in the ($\Sigma$, $T_{\rm c}$) plane
for annuli at $2\times 10^{10}$cm (left panels)
and $6\times 10^{10}$cm (right panels) with both direct and indirect
irradiation
during the decay phase from the outburst maximum
to the quiescent minimum state.
In the upper (a) and (c)
the open and filled circles represent epochs when the zone
is and is not subject to direct irradiation, respectively.
The outer region, at $6\times 10^{10}$cm, receives no direct irradiation
due to the blocking by the inner disk and hence the shadowing
is in effect from the outburst maximum until the disk returns to the
cool quiescent state (see figure 4).
The numbered epochs correspond to times after maximum in days
as follows: 1 -- 0.94; 2 -- 3.5; 3 -- 4.7; 4 -- 6.9; 5 -- 14;
6 -- 15; 7 -- 21; 8 -- 41.6; 9 -- 41.9; 10 -- 44; 11 -- 53; 12 -- 54;
13 -- 55; 14 -- 57.2; 15 -- 57.4; 16 -- 57.5; 17 -- 70; 18 -- 88.2;
19 -- 88.4; 20 -- 94; 21 -- 130; 22 -- 211.
In the lower (b) and (d)
the trajectory of the annulus evolution is illustrated
for three cases: non-irradiated (dotted line), directly irradiated only
(dashed), and indirectly irradiated only (solid line) models.}
\end{Fv}

Figure 8 presents the actual track of the evolution of the
surface density and mid-plane temperature $T_{\rm c}$, the so-called
thermal limit cycle (for reviews, see Osaki 1989; Cannizzo 1993)
for various models.
The top panels in figure 8 present the evolution of the region
around $1-2\times10^{10}$ cm (panel a) and around
$4-6\times10^{10}$ cm (panel c) for the case with both direct
and indirect irradiation. The open
circles show when the region is subject to direct irradiation and
the closed circles show when the region is shadowed from direct irradiation
at the indicated times.
For the same two regions, the lower panels in figure 8
present models corresponding to no irradiation (long dash),
direct irradiation only (short dash), and
indirect irradiation only (dotted line).
The model with only indirect irradiation shows very similar evolution
to the full model.
     There are three distinctive stages to the full model:

\noindent
(1) In the region from $1-2\times10^{10}$ cm,
the temperature stops declining and then
climbs back up to the hot state again (figure 8a, time steps 9$-$14).

\noindent
(2) On the second decline from the hot state
the disk temperature {\it stagnates} at
$\log T_{\rm c}$ $\sim$ 3.8$-$4.1 (figure 8a, time steps 16$-$19)
in the partially ionized region (see figure 7 and subsection 4.2), 
prior to the final
downward transition to the cool disk.  We call this phase
an ``ionization stagnation."

\noindent
(3) In the outer region, $\sim4-6\times10^{10}$ cm,
the disk shows a similar stagnation, but it occurs at
$\log T_{\rm c}$ $\sim$ 3.4$-$3.6 (figure 8c, time steps 3$-$7)
in the molecular opacity-dominated region (see figure 7 and section 4.2).
We call this phase  ``molecular stagnation."

Figure 8 shows that the stagnation behavior on the decline is not present
in the non-irradiated or only directly irradiated models.
Even with modest indirect irradiation this stagnation behavior
can affect the outburst and cooling of the disk.

\subsection{ Stagnation Phenomena}

The disk is in thermal equilibrium when the viscous heating, 
$Q^+_{\rm vis}$ [equation (\ref{qvis})], 
balances the radiative cooling, $Q_{\rm rad}^-$ [equation (\ref{qrad})], 
or $Q^+_{\rm vis}$~=~$Q^-_{\rm rad}$.
In the thin disk approximation (e.g., Shakura, Sunyaev 1973), the
emergent heat flux radiated vertically in the plane-parallel approximation is
given by $F(z)$ = $4{\sigma}{T_{\rm c}}^4/3{\tau}$.
Since $\tau$ = ${\Sigma}{\kappa}$ for the Rosseland mean opacity $\kappa$ 
and the surface density is $\Sigma$ = $2{\rho}H$ for density $\rho$ and disk
height $H$, one can write
\begin{equation}
\label{teff}
     T_{\rm eff}^4~=~{ {8{T_{\rm c}}^4}\over {3{\kappa}{\Sigma}} }.  
\end{equation}
Since the sound speed is
$C_{\rm s}$ = $({{\cal R}{T_{\rm c}}/{\mu}})^{1/2}$, 
where $\cal R$ is the gas constant and 
$\mu$ is the molecular weight, and H = $C_{\rm s}$/${\Omega}$,
the energy equation in thermal equilibrium becomes
\begin{equation}
\label{qq}
     {3\over 2}\alpha\Sigma
      \left({{\cal R}T_{\rm c}\over \mu}\right){\Omega}~=~
     {3\over 2}{\alpha}{\Omega}{P\over {\rho}}~=~
     {{4ac{T_{\rm c}}^4}\over {3{\kappa}{\Sigma}}},
\end{equation}
where $\sigma$ $\equiv$ ac/4 
and $\kappa$ = $\kappa$($T_{\rm c}$, $\rho_{\rm c}$).
Note that, in the disk-instability model which we present here, 
$\alpha$ is also a function of $T_{\rm c}$ [see equation (\ref{alpha})]. 
In various physical regimes, the opacity can be expressed as 
\begin{equation}
     {\kappa}~=~{\kappa}_0{\rho_{\rm c}}^m{T_{\rm c}}^n,
\end{equation}
where ${\kappa}_0$ is a constant and $m$ $\geq$ 0, but $n$ can be
negative or positive (see figure 1).
     
The possibility of stagnation in a partially ionized region was first
pointed out by Meyer-Hofmeister (1987) and elaborated by Mineshige (1988).
For the case of a departure from thermal equilibrium, the
energy equation, equation (\ref{energy}), can be written as
\begin{equation}
     {C_{\rm p}\Sigma\over 2}{\partial T_{\rm c}\over {\partial t}} ~=~
	Q^+ - Q^-,
\end{equation}
where $C_{\rm p}$ is the specific heat at constant pressure,
and $Q^+$ and $Q^-$ are the total heating and cooling rates.
The rate of change of the temperature can be decreased
by increasing the specific heat, or by decreasing the departure 
from thermal equilibrium, Q$^+\sim$ Q$^-$, or a combination
of both.  The essence of the higher temperature stagnation is that the 
specific heat increases in the regions of partial ionization and
if the structure is not too far from thermal equilibrium, the thermal
time scale is increased, leading to a ``metastable" intermediate
temperature state.
In the rise, the partially ionized region is near the ``knee"
in the instability curve, where Q$^+\sim$ Q$^-$ and the effect
can be pronounced.  

Mineshige (1988) showed that the onset of the
heating can consist of a ``warming wave" which raises
the disk from the cold state to this intermediate metastable state.
This state is ultimately unstable and a full heating wave causing
the transition to the hot state finally ensues.
In the decline, the partial ionization state is further from
the equilibrium curve, and although the specific heat increases, 
the cooling term, Q$^-$, dominates the heating term, and there is
little effect of the specific heat as the 
disk drops into the cool state.  As a result,
stagnation has not been manifested in the decline in previous
disk-instability models.  As the current models illustrate, 
irradiation can enhance the heating term, Q$^+$, in the decline,
thus restoring an approximate balance, Q$^+\sim$ Q$^-$.  In 
this situation, the effect of an increase of C$_{\rm p}$ can be
seen as  ``stagnation" in the decline as well. 
This can cause the disk to linger in the intermediate-temperature state.  
It eventually makes a downward transition into the cool state.

At $\log T_{\rm c}$ $\lta$ 3.5$-$3.6, 
the molecular absorbers become important and
at $\log T_{\rm c}$ $\sim$ 3.4, ${\rm {H_2O}}$ dominates.
The low-temperature opacity is dominated by grains
as grain condensation begins at
$\log T_{\rm c}$ $\lta$ 3.25.
These various effects change the slope of the opacity
at about $\log T_{\rm c}$ = 3.25, 3.4 and 3.5$-$3.6 (figure 1).
As a result, when the disk passes through $\log T_{\rm c}$ = 3.6,
the opacity suddenly {\it increases} with {\it decreasing} temperature
[see equation (\ref{teff})].
As a result, the thermal time scale for the downward transition will be
increased.

\section{Discussion and Conclusions}

The purpose of this paper was to investigate some of the basic
time-dependent effects of irradiation.
We show that even relatively mild irradiation from the inner disk
can affect the overall disk evolution.
Mild irradiation can affect the limit cycle
by causing portions of the disk to linger in the 
intermediate-temperature metastable ``stagnation state."
 In contrast to the non-irradiated disk-instability models, the
irradiated models show a wave of cooling to the stagnation state
that propagates inward in advance of cooling to the low,
molecular-dominated state.  
  It is not clear that this is simply a result of heating
of the disk, since the non-linear interplay of the zones seems
to have an important role.  The issue of how a particular
portion of the disk responds to irradiation by tracking along
the metastable state 
is a rather subtle thing involving complex interactions
among the different portions of the disk. 
Due to shadowing, the effect of direct irradiation terminates as each
annulus of the disk enters the stagnation state.
The net result of the irradiation is a slower decay of the light curve.

When an outburst is initiated, the outer disk is exposed
to direct irradiation, but that radiation is feeble until
the heating wave reaches the inner disk.  Soon thereafter
the outer disk is shadowed by a swelling of the middle of the disk.
Direct irradiation does not play an important
role in the behavior of the outer disk either during 
and after the maximum of the outburst,
but can affect the middle of the disk during the outburst.
Since it is by assumption not influenced by shadowing,
the indirect irradiation affects the overall disk evolution,
including the outer disk, throughout its evolution.

As stated in the introduction, this is the first study of time-dependent
irradiation of black-hole accretion disks.  For this reason and technical
issues related to code instability, we have presented models with
rather crude zoning.  These models are a major step beyond the one-zone
models of Tuchman et al.(1990), but only a step toward the goal
of self-consisent,  resolved irradiated models.  Beside the
rather crude zoning, other restrictions are rather simplified models for
the irradiation that make no attempt to do realistic radiative transfer
and, of course, continued uncertainty about the physics of disk viscosity
and the role of disk evaporation.
In the context of these general limitations and uncertainties, these
models do give some insight into the issues of time-dependent irradiation.
They incorporate sufficient physics to capture the non-monotic nature
of the disk profile, a critical issue in considering shadowing.
Given this context, there are some qualitative aspects of these models
that are reasonably trustworthy:
\begin{itemize}
\item The direct irradiation of the disk by inner layers where the
    soft X-ray flux is generated is found to have only a small effect
    on the outer disk because of shadowing.
\item Mild indirect irradiation that flattens, but otherwise does not affect 
    the light curve substantially, still has interesting non-linear effects 
    on the structure of the disk as heating and cooling waves propagate.
\end{itemize}
Any quantitative results (e.g., front propagation velocity, the detailed
shape of the light curves, etc) may change when fine-mesh calculations 
are performed.  Specifically, the calculated reflare-like optical events may
or may not be a calculation artifact.  We need fine-mesh calculations to 
confirm the reflare-like behavior of the irradiated disk, but the
basic phenomenon of radiation-induced stagnation allowing a temporary
return to the hot state is physically plausible. It is not clear how
such a phenomenon could lead to a modulation of the soft X-rays,
a defining feature of the observed reflares. 
   
Saito (1989) considered the case of a central star 
that illuminated the disk
using a simple one-zone model for the vertical disk structure, and
thus did not properly consider the non-(thermal)-equilibrium states
of the disk.
The irradiation from a central source is
larger than what we consider here by a factor of $H/R$
(cf. subsection 2.1, van Paradijs 1996).
In addition, Saito neglected a term equivalent to [$d\ln H/d\ln R-1$] =
$(R/H)$[$d{(H/R)}^2/dR$] [see equation (3)],
that measures the shape of the disk profile.  He took the angle of 
incidence of the radiation, $\sim H/R$, and the flux generated by
accretion to be constant. Despite these differences and approximations,
Saito's results are qualitatively similar to ours.  With a large
accretion rate ($10^{17}$ g s$^{-1}$) and the various approximations
that enhance the irradiation, Saito found that the inner disk
was always maintained in a permanently hot state.  
The outer
disk, however, underwent a thermal instability that began in the
outer part of the disk. Saito notes that prior to the outburst
the outer disk is in the shadow of the inner disk.

The disk instability due to the ionization of hydrogen and helium
remains the most plausible cause of the outburst of the black-hole 
candidate X-ray novae.  For the orbital periods and mass-transfer 
rates inferred from observations, the disk is predicted to
be unstable.  A steady state is very unlikely.  Because the 
direct component of irradiation comes from a surface of the disk
that is nearly in the plane of the outer disk, and hence presents
a small solid angle, disks around black holes receive relatively
little direct irradiation (Cannizzo 1994; King et al. 1997;
Cannizzo 1998; Dubus et al. 1999). Gradients in opacity, 
the viscosity, the effect of the inner boundary condition,
and of time-dependent structure, all serve to diminish
the direct irradiation even further.   Whether a source
of ``indirect" irradiation as we have modeled it here can
have a more severe effect, including stabilizing the disk
(van Paradijs 1996), requires further investigation.

The value of the indirect irradiation parameter, $C_{\rm X}\sim10^{-4}$,
was chosen in conjunction with other parameters in these
models to reproduce the approximate slope of the optical light
curve of A0620--00.  We find that this level of irradiation
does not significantly affect the luminosity at maximum, but
does make the disk brighter by about 0.5 mag at about 50 d
after maximum.  Cannizzo (1998) has estimated that A0620--00
might be brighter by about 1 mag at maximum light
than his standard non-irradiated models.  This difference
may be within the model uncertainties, but, if ascribed to
irradiation, it could be accounted for with a modest
increase in $C_{\rm X}$.
    
We have shown that even modest levels of irradiation can
influence the thermodynamics of the disk and affect its
evolution.  It is very important to consider indirect
as well as direct irradiation of the disk, since the effect
of the latter is severely constrained by shadowing.
                                                            
Finally, we can consider how our results might change if the
quiescent disk has a hole in the middle because of gas evaporation
and resultant advection-dominated flow.  Such a hole will
change, for example, the quiescent surface-density distribution,
thereby affecting the ignition time and radius (Hameury et al. 1997).
Also, the disk should expand during quiescence
because of continuous input of angular momentum by the incoming 
stream, while the mass increase will be modest (Mineshige et al. 1998).
Once an outburst occurs, however, the entire disk evolution will
be totally controlled by the propagation of the heating front, 
except at the very early rise phase, so that
our conclusions will not change significantly.

\par
\vspace{1pc} \par
We are grateful to John Cannizzo, Wan Chen, Neil Gehrels, 
Robert Hjellming,  Shunji Kitamoto, Masaaki Kusunose, Jean-Pierre Lasota, 
Don Lamb, Jeff McClintock, Yoji Osaki,
Edward Robinson, Divas Sanwal, Chris Shrader, Ethan Vishniac,
and the GRO BATSE NASA/MSFC team, especially to Alan Harmon, 
William Paciesas, and Jan van Paradijs for fruitful discussions
and to the anonymous referee for comments that helped to clarify
the manuscript.
We acknowledge the optical light curves of A0620--00 sent by Ron Webbink
and the optical light curves of Nova Sco 1994 sent
by Jerome Orosz and Charles Bailyn.
S.-W. K. gives special thanks to 
Seok-Jae Lee, Gei-Youb Lim, Myung-Shin Chung, Kyong-Sei Lee, and
Kwang-Souk Sim of the Nuclear Physics Lab of Korea University for their
hospitality and help with computation 
and plots during his stay in Seoul, Korea.
S.-W. K. thanks to Hwankyung Sung, Won-Kee Park, and Hak Min Kim
for helping to prepare figures 
and Seung Soo Hong for his encouragement during the first year of
the KRF fellowship.
Numerical calculations were performed at the High Performance Computing 
Facility (HPCF) of the University of Texas at Austin.
This research was supported in part by NSF Grants AST 9115143, AST-9218035 
and AST-9528110, NASA Grant NAGW 2975, a grant 
from the Texas Advanced Research Program (JCW),
and the Grants-in Aid of the
Ministry of Education, Science, and Culture of Japan
(10640228, SM).
S.-W. K. wishes to acknowledge the financial support of the Korea
Research Foundation made in the program Year 1997$-$1998 and Year 1998$-$1999.
  
\section*{References}
\small

\re
Alexander D. R., Johnson H. R.,  Rypma R. L. 1983, ApJ 272, 773

\re
Cannizzo J. K. 1993, in Accretion Disks in Compact Stellar Systems,
     ed J. C. Wheeler (World Scientific: River Edge, NJ) p6

\re
Cannizzo J. K. 1994, ApJ 435, 389 

\re
Cannizzo J. K. 1998, ApJ 494, 366 

\re
Cannizzo J. K., Chen W.,  Livio M. 1995, ApJ 454, 880

\re
Cannizzo J. K., Ghosh P.,  Wheeler J. C. 1982, ApJ 260, L83

\re
Cannizzo J. K., Shafter A. W.,  Wheeler J. C. 1988, ApJ 333, 227

\re
Cannizzo J. K.,  Wheeler J. C. 1984, ApJS 55, 367

\re
Chen W., Shrader C. R.,  Livio M. 1997, ApJ 491, 312

\re
Cox A. N.,  Stewart J. N. 1970, ApJS 19, 261

\re
Cox A. N.,  Tabor J. E. 1976, ApJS 31, 271
                                                           
\re
de Jong J. A., van Paradijs J., Augusteijn T. 1996, A\&A 314, 484

\re
Dubus G., Lasota J.-P., Hameury J.-M., Charles P. 1999, 
MNRAS 303, 139

\re
Frank J., King A.,  Raine D. 1992, Accretion Power in
     Astrophysics, 2nd ed (Cambridge University Press, Cambridge) ch4

\re
Fukue J. 1992, PASJ 44, 663

\re
Hameury J.-M., Lasota J.-P., McClintock J.E., Narayan R. 1997, ApJ 489, 234

\re
Kim S.-W., Wheeler J. C.,  Mineshige S. 1992, ApJ 384, 269 
     and ApJ 399, 330 (erratum)

\re
Kim S.-W., Wheeler J. C.,  Mineshige S. 1994, in The Evolution
of X-ray Binaries, AIP 308, ed S. S. Holt, C. S. Day
(AIP, New York) p213 

\re
Kim S.-W., Wheeler J. C.,  Mineshige S. 1996, in Basic Physics of
Accretion Disks, ed S. Kato, S. Inagaki, J. Fukue, S. Mineshige
 (Gordon and Breach, New York) p171 


\re
King A. R., Kolb U.,  Szuszkiewicz E. 1997, ApJ 488, 89

\re
Kusaka T., Nakano T., Hayashi C. 1970, Prog.Theor.Phys. 44, 1580

\re
Kuulkers E. 1998, New Astronomy Rev. 41, 1

\re
Ludwig K., Meyer-Hofmeister E., Ritter H. 1994, A\&A 290, 473

\re
Lund N. 1993, A\&AS 97, 289

\re
Marsh T. R., Robinson E. L., Wood J. H. 1994, MNRAS 266, 137

\re
McClintock J. E., Horne K.,  Remillard R. A. 1995, ApJ 442, 358

\re
McClintock J. E.,  Remillard R. A. 1986, ApJ 308, 110

\re
Meyer F., Meyer-Hofmeister E. 1984, A\&A 140, L35

\re
Meyer-Hofmeister E. 1987, A\&A 175, 113
                                                       
\re
Mineshige S. 1986, PASJ 38, 831 

\re
Mineshige S. 1988, A\&A 190, 72

\re
Mineshige S. 1994, ApJ 431, L99

\re
Mineshige S., Liu B., Meyer F., Meyer-Hofmeister E. 1998, PASJ 50, L5

\re
Mineshige S.,  Osaki Y. 1983, PASJ 35, 377 

\re
Mineshige S.,  Osaki Y. 1985, PASJ 37,1 

\re
Mineshige S., Kim S.-W.,  Wheeler, J. C. 1990a, ApJ 358, L5

\re
Mineshige S., Tuchman Y.,  Wheeler J. C. 1990b, ApJ 359, 176

\re
Mineshige S.,  Wheeler J. C. 1989, ApJ 343, 241

\re
Mineshige S.,  Wood J. H. 1989, MNRAS 241, 259

\re
Osaki Y. 1989, in Theory of Accretion Disks, ed F. Meyer,
W.J. Duschl, J. Frank, E. Meyer-Hofmeister
     (Kluwer Academic Publishers, Dordrecht) p183
                                                       
\re
Paczy{\'n}ski B. 1977, ApJ 216, 822

\re
Saito N. 1989, PASJ 41, 1173
 
\re
Shakura N. I.,  Sunyaev R. A. 1973, A\&A 24, 337

\re
Tuchman Y., Mineshige S.,  Wheeler J. C. 1990, ApJ 359, 164 

\re
van Paradijs J. 1996, ApJ 464, L139
                                              
\re
Webbink R. F. 1978, A provisional optical light curve of the X-ray
     recurrent nova: V616 Monocerotis = A0620--00 
    (University of Illinois, Urbana-Champaign)
 
\re
Wheeler J. C. 1997, in Relativistic Astrophysics, 
ed B. Jones, D. Markovic (Cambridge University Press, Cambridge)  p211

\re
Wheeler J. C. 1999, in Disk Instabilities in Close Binary Systems:
25 Years of the Disk-instability Model, ed S. Mineshige,
J.C. Wheeler (Universal Academy Press, Inc., Tokyo)  p31

\re
Wheeler J. C., Kim S.-W., Moscoso M., Kusunose M., Mineshige M. 
1996, in Basic Physics of Accretion Disks, ed S. Kato, S. Inagaki, J. Fukue, S. Mineshige
(Gordon and Breach, New York) p127   

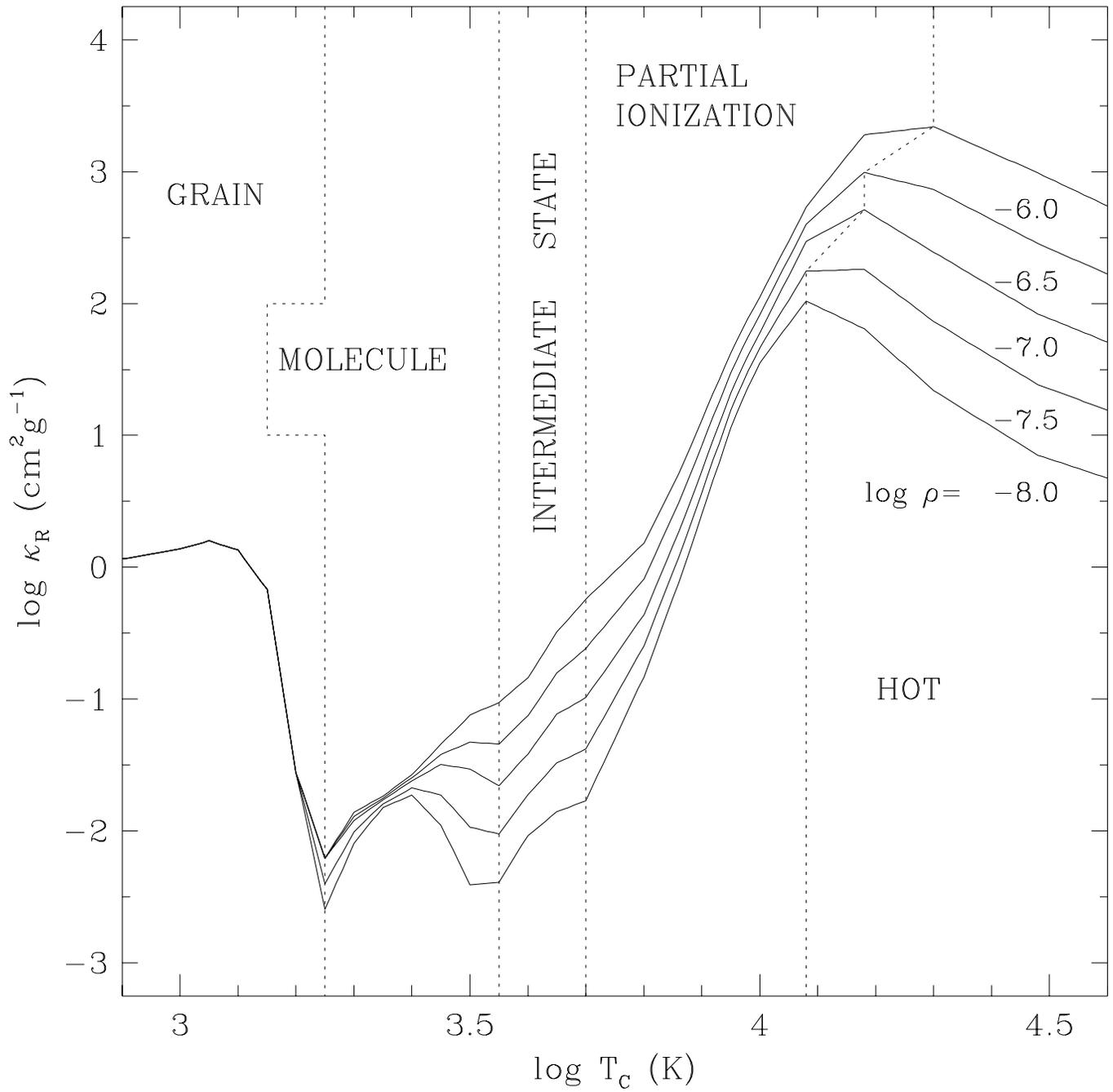
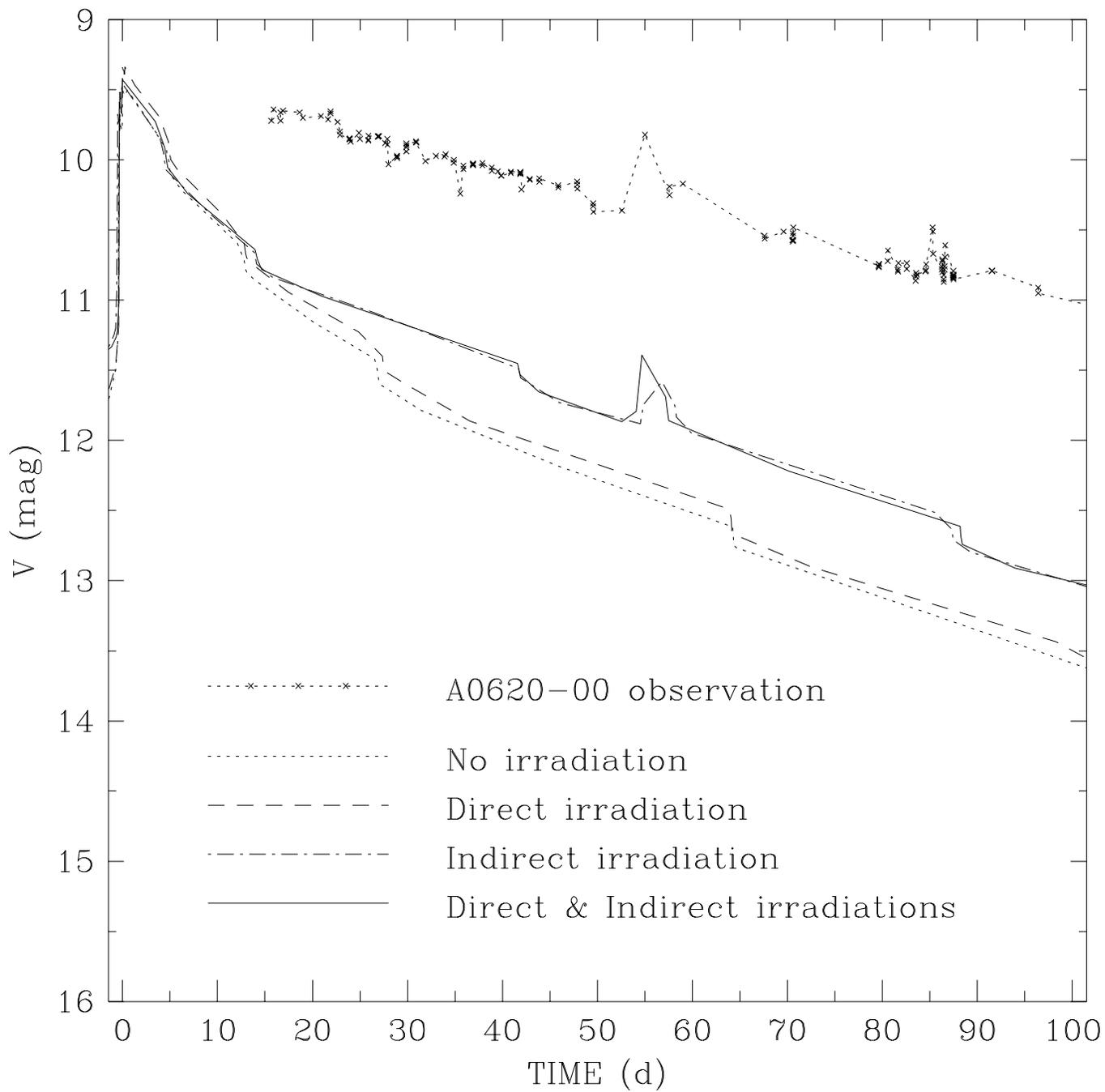
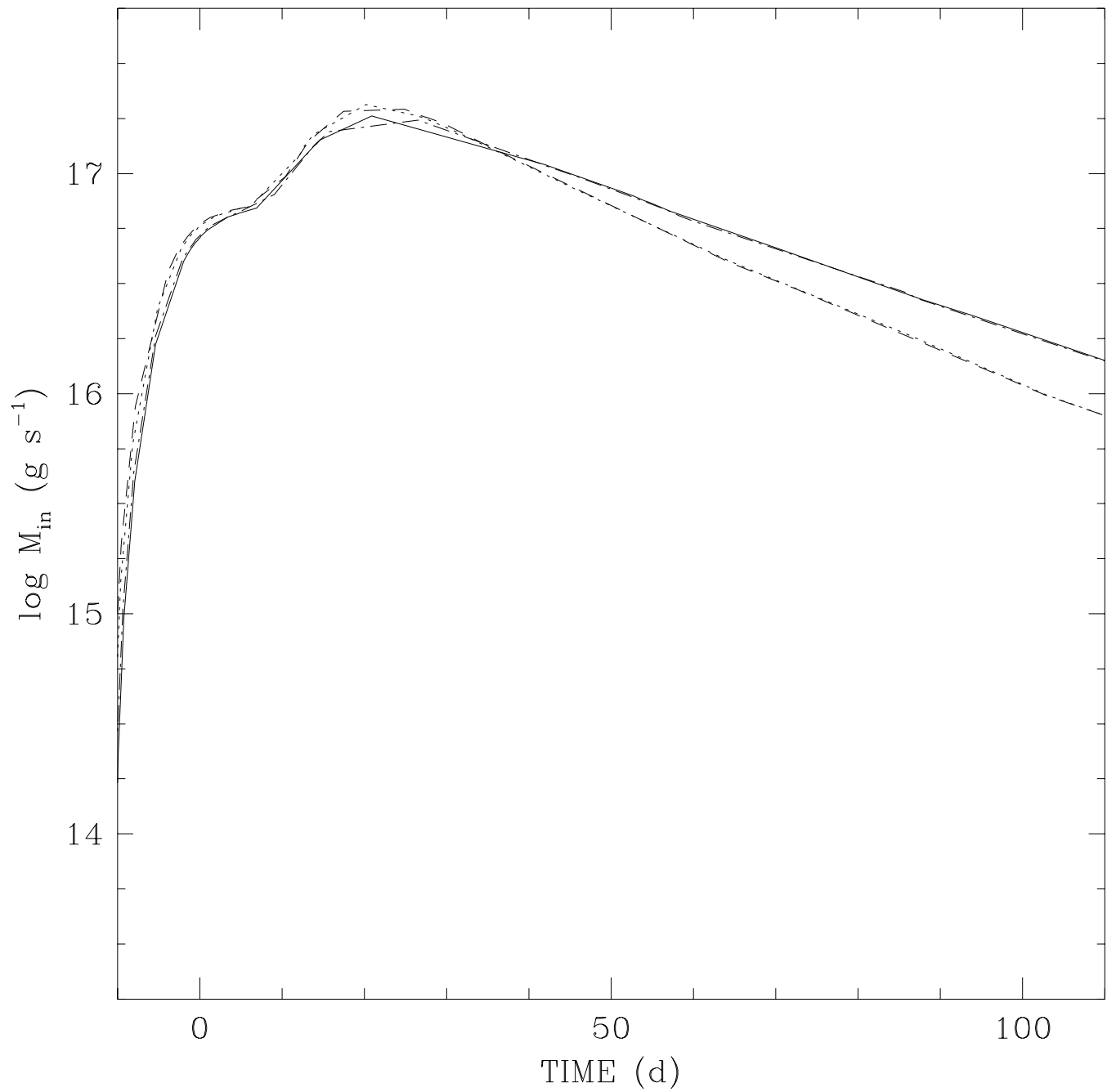
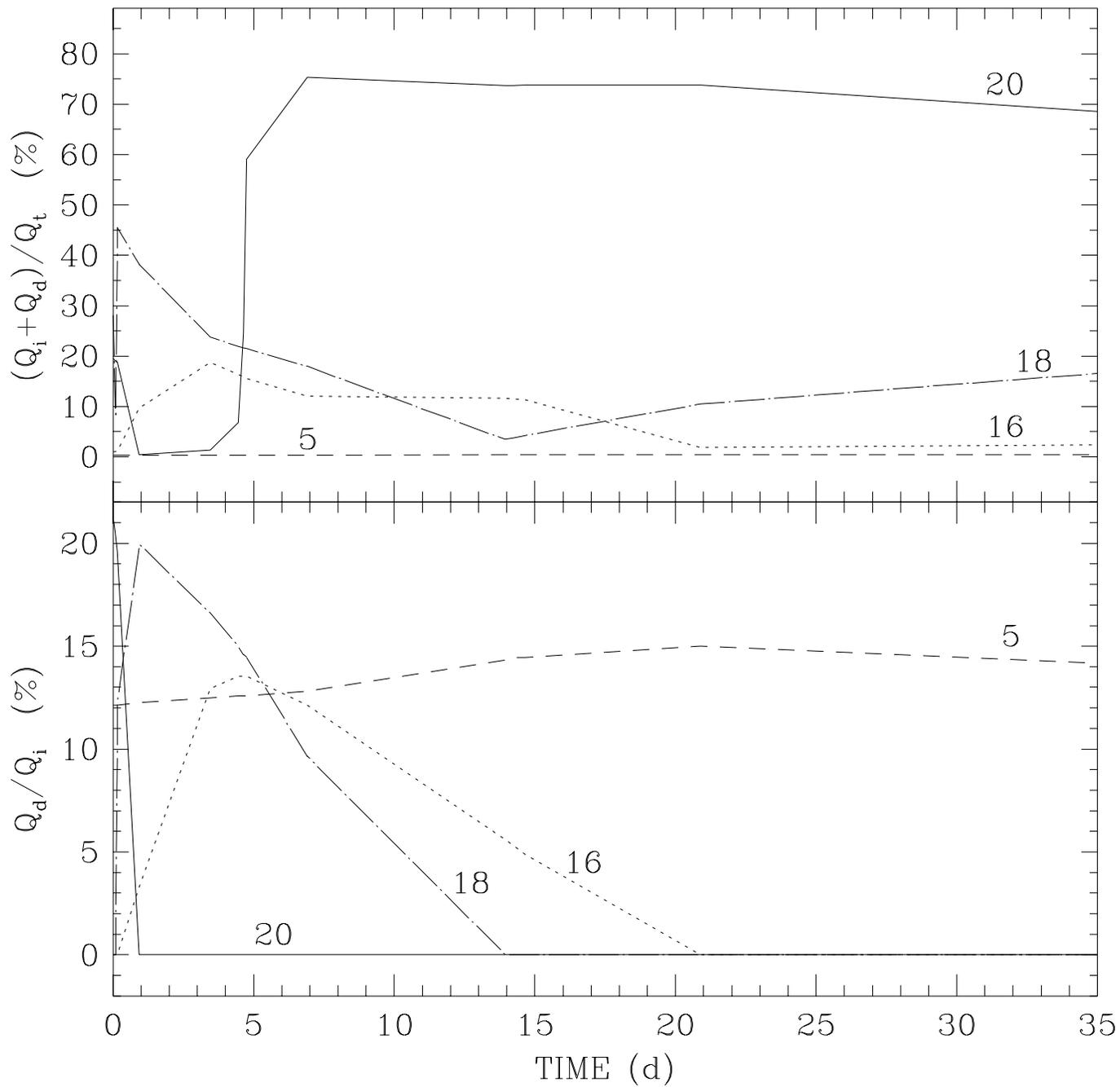
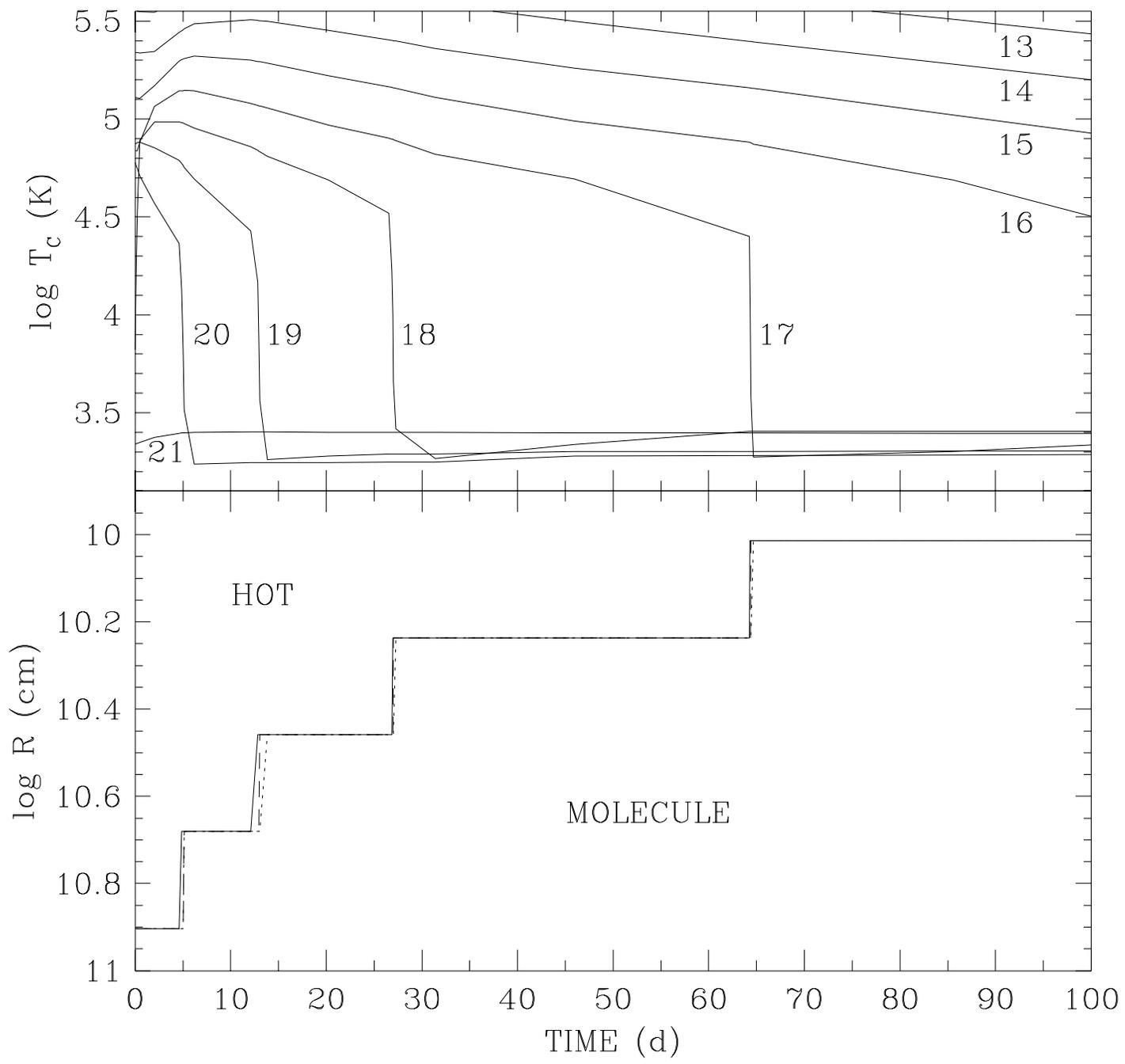
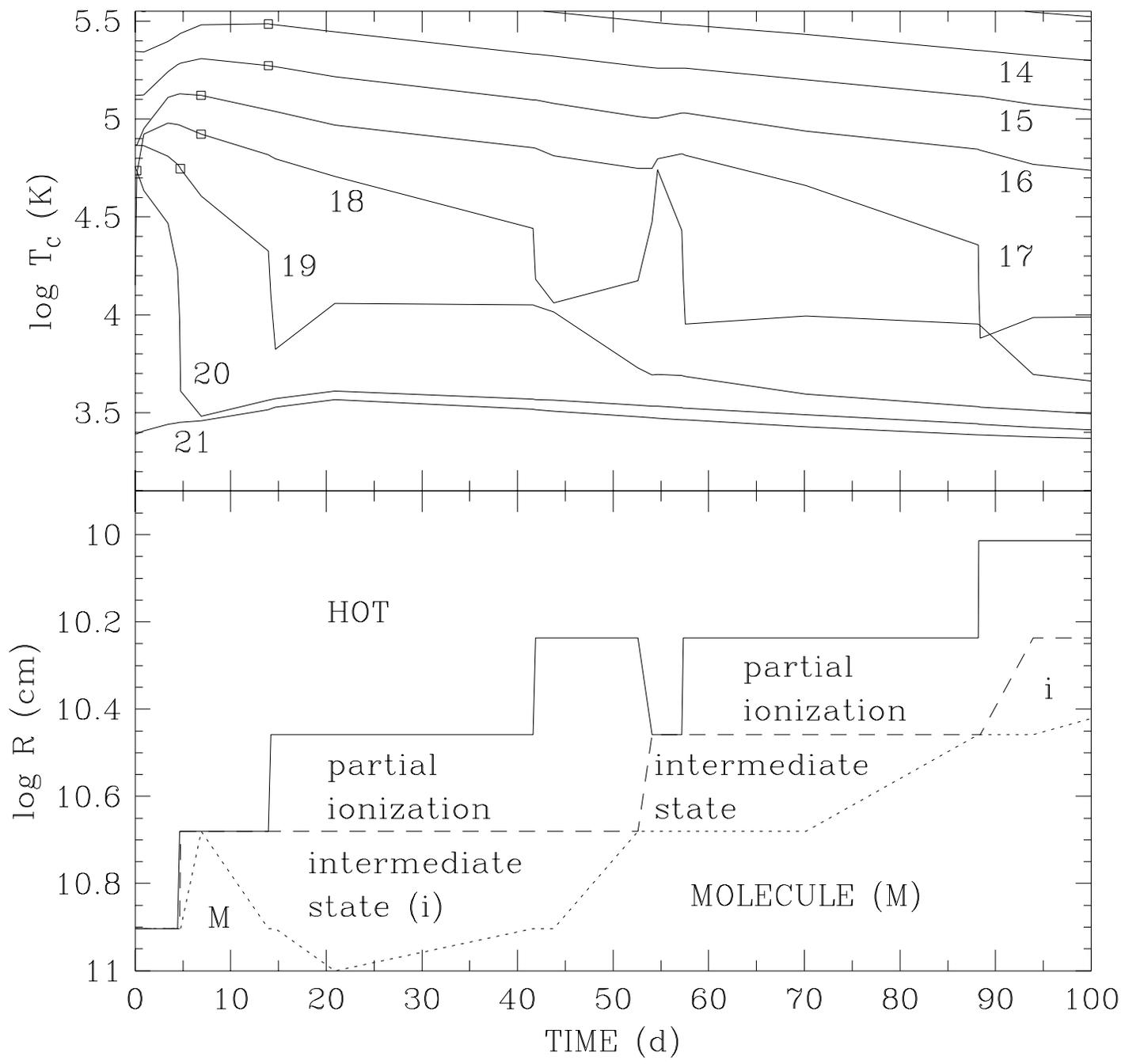
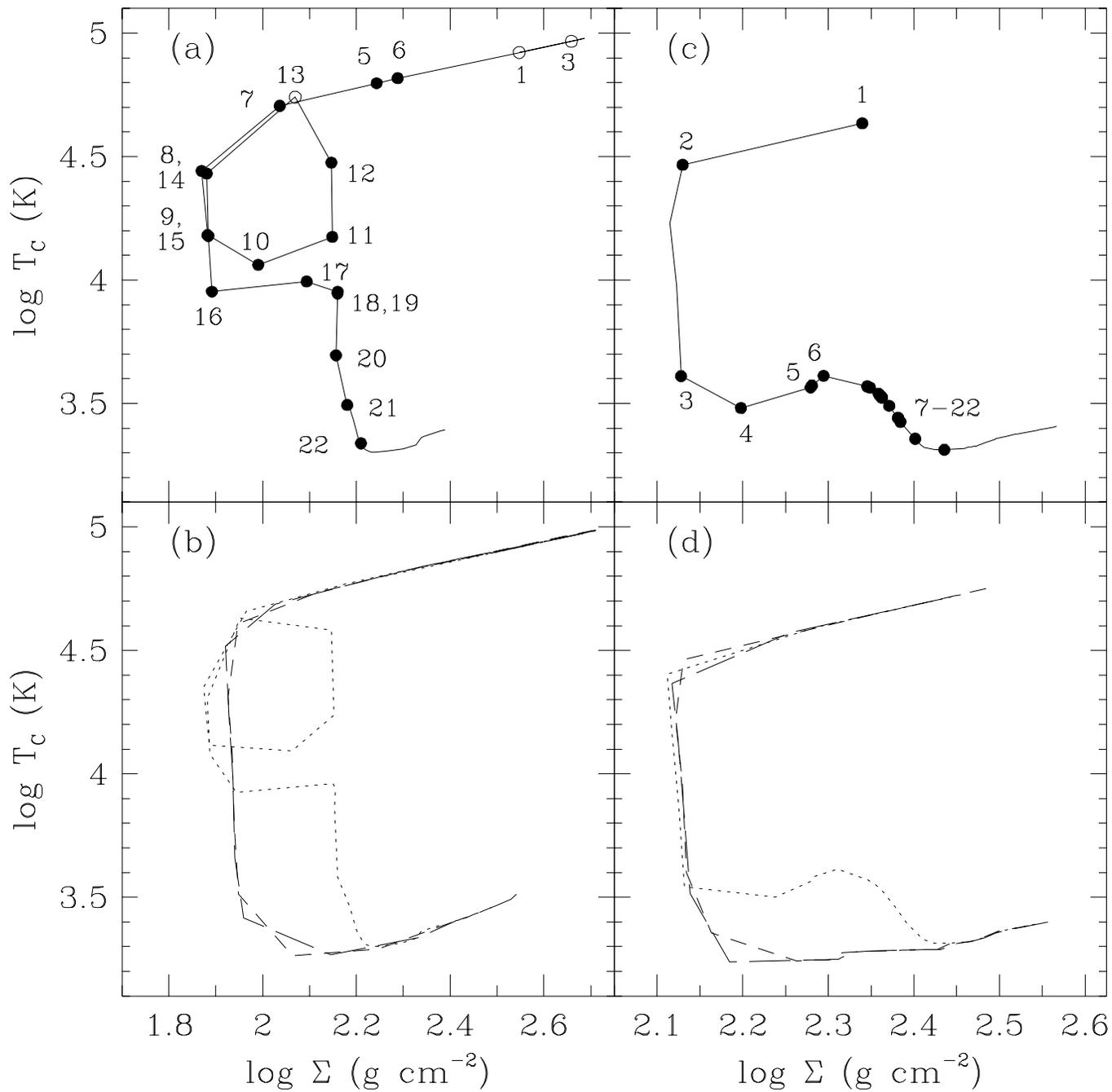
\end{document}